\documentclass[aps, prl,english,twocolumn,showpacs,10pt, longbibliography, superscriptaddress]{revtex4-2} 
\usepackage{amsmath,bm}
\usepackage{mathtools}
\usepackage{commath}
\usepackage{amssymb}
\usepackage{graphicx}
\usepackage{babel}
\usepackage{cases}
\usepackage[colorlinks=true, citecolor=blue, anchorcolor=green]{hyperref}
\hypersetup{linkcolor=red}
\usepackage{natbib}
\usepackage{floatrow}
\usepackage{dblfloatfix}
\usepackage{xcolor}
\usepackage{braket}

\makeatletter

\newcommand{\Rmnum}[1]{\expandafter\@slowromancap\romannumeral #1@}

\newcommand{\eff}{\text{eff}}

\newcommand{\be}{\begin{equation}}
\newcommand{\ee}{\end{equation}}
\newcommand{\ba}{\begin{aligned}}
\newcommand{\ea}{\end{aligned}}

\makeatother

\begin{document}
\title{Singular band Induced by Long-Range Interaction
Enables Unsplit Spreading of Localized Excitations}
\author{Jian-Feng Wu}
\author{Yi Huang}
\affiliation{Institute of Physics, Chinese Academy of Sciences, Beijing 100190, China}
\affiliation{School of Physical Sciences, University of Chinese Academy of Sciences, Beijing 100049, China}
\author{Yu-Xiang Zhang}
\email{iyxz@iphy.ac.cn}
\affiliation{Institute of Physics, Chinese Academy of Sciences, Beijing 100190, China}
\affiliation{School of Physical Sciences, University of Chinese Academy of Sciences, Beijing 100049, China}
\date{\today}

\begin{abstract}
    In conventional lattice models, the dispersion relation $\omega(k)$ is assumed to be a smooth function which is periodic over the first 
    Brillouin Zone. However, in subwavelength atom arrays 
    the dispersion of the 
    light-mediated long-range interaction is singular at the light cone. This observation prompts us to ask what effect arises from such band singularity.   
    Here we demonstrate that, due to the topology of smooth functions defined over the periodic Brillouin zone, smoothness
    implies the splitting of an initially localized excitation into counter-propagating wave packets. Consequently, unsplit spreading can occur only when $\omega(k)$ develops singular features, precisely what long-range interactions enable. We identify unsplit spreading in 1D toy 
    tight-bounding models and the realistic models of 1D and 2D subwavelength atomic arrays.
    Our work establishes unsplit spreading as an experimentally accessible, smoking-gun signature of singular band structure.
\end{abstract}

\maketitle

A persistent theme in the study of long-range interacting systems is to 
delineate how models with couplings decaying as $1/r^\alpha$ 
(where $r$ is the distance and $\alpha>0$) depart from their short-range 
counterparts~\cite{Defenu:2023aa,Mattes:2025aa,Eisert:2013aa,Liu:2019aa,Defenu:2024aa,Koffel:2012aa,Schachenmayer:2013aa,Lerose:2020aa}.  
Such long-range terms are typically introduced in the free part of the 
Hamiltonian, e.g., as hopping amplitudes in tight-binding models.  
As long as the system is not strongly correlated, i.e., when the interaction can 
be treated as a perturbation, the free Hamiltonian sets the tone for all 
essential dynamical features.  
Even at this level, the contrast between short- and long-range models is 
significant.  
In the regime of \emph{strong long-range interactions}~\cite{Defenu:2023aa}, 
where $\alpha<d$ (with $d$ the spatial dimension), the dispersion relation 
$\omega(k)$ develops a point spectrum, discrete eigenvalues instead of 
continuous bands~\cite{Defenu:2021aa}.  
This property underlies several anomalous phenomena, including the absence of 
equilibration~\cite{Kastner:2011aa} and the persistence of finite Poincar\'{e} recurrence times~\cite{Defenu:2021aa}.  
We note that even for $\alpha\geq d$, the dispersion can still exhibit 
singularities, albeit less drastic than those in the $\alpha<d$ case.  
By contrast, short-range tight-binding models always yield smooth dispersion relations.

Singular band structures have also been widely noticed, though seldom studied systematically, in the context of resonant dipole–dipole interactions, which constitute a fundamental interaction rather than one realized in delicately protected quantum simulators.  
In free space, resonant dipole-dipole interactions is a mixture of different $1/r^\alpha$ terms with $\alpha=1,2$, and $3$~\cite{Lehmberg:1970aa,Lehmberg:1970ab}.  
The interaction is established by exchanging photons resonant with a specific transition frequency $\omega_0$, corresponding to the light cone at photon momentum $k_0=\omega_0/c$ ($c$ is the speed of light).  
To see how a singular band structure emerges, consider a one-dimensional (1D) atomic array with lattice constant $a$.  
The light cone is enclosed by the first Brillouin zone when $\pi/a>k_0$, i.e., when $a$ is smaller than half the resonant wavelength, hence the term \emph{subwavelength atom array}~\cite{Facchinetti:2016aa,Shahmoon:2017aa,Plankensteiner2017,Guimond:2019aa,Bekenstein:2020aa,Rui:2020aa}.  
The physics of Bloch modes with $|k|<k_0$ and $|k|>k_0$ is then fundamentally different: excitations in the former are radiative, whereas spontaneous emission from the latter is forbidden, a phenomenon known as subradiance~\cite{Dicke1954,Bienaime2012,Guerin2016,Jenkins2017,Weiss:2018aa,Yan:2023aa}.  
Correspondingly, the dispersion relation becomes singular exactly at the boundary $|k|=k_0$.

This observation raises a direct question: what observable consequences follow from the singularity?  
The question is less trivial than it may first appear.  
Although long-range resonant dipole-dipole interaction is the microscopic mechanism behind subradiance~\cite{Lehmberg:1970aa,Lehmberg:1970ab,Dung2002}, 
short-range toy models can nevertheless reproduce many key quantitative 
features, such as the power-law scaling of subradiant decay 
rates~\cite{Tsoi:2008aa,Asenjo-Garcia2017,Albrecht2018,Kornovan:2019aa} and the 
antisymmetric wavefunctions of multiply excited subradiant states~\cite{Asenjo-Garcia2017,Albrecht2018}.  
This is possible because these features require only local knowledge of 
$\omega(k)$ near specific $k$-points, which short-range models can faithfully 
capture~\cite{Zhang:2019aa,Zhang:2020ab,Zhang:2022aa}.  
Moreover, long-range interactions sometimes behave effectively like
even noninteracting systems~\cite{Mattes:2025aa}.

\begin{figure}[t]
    \centering
    \includegraphics[width=0.9\textwidth]{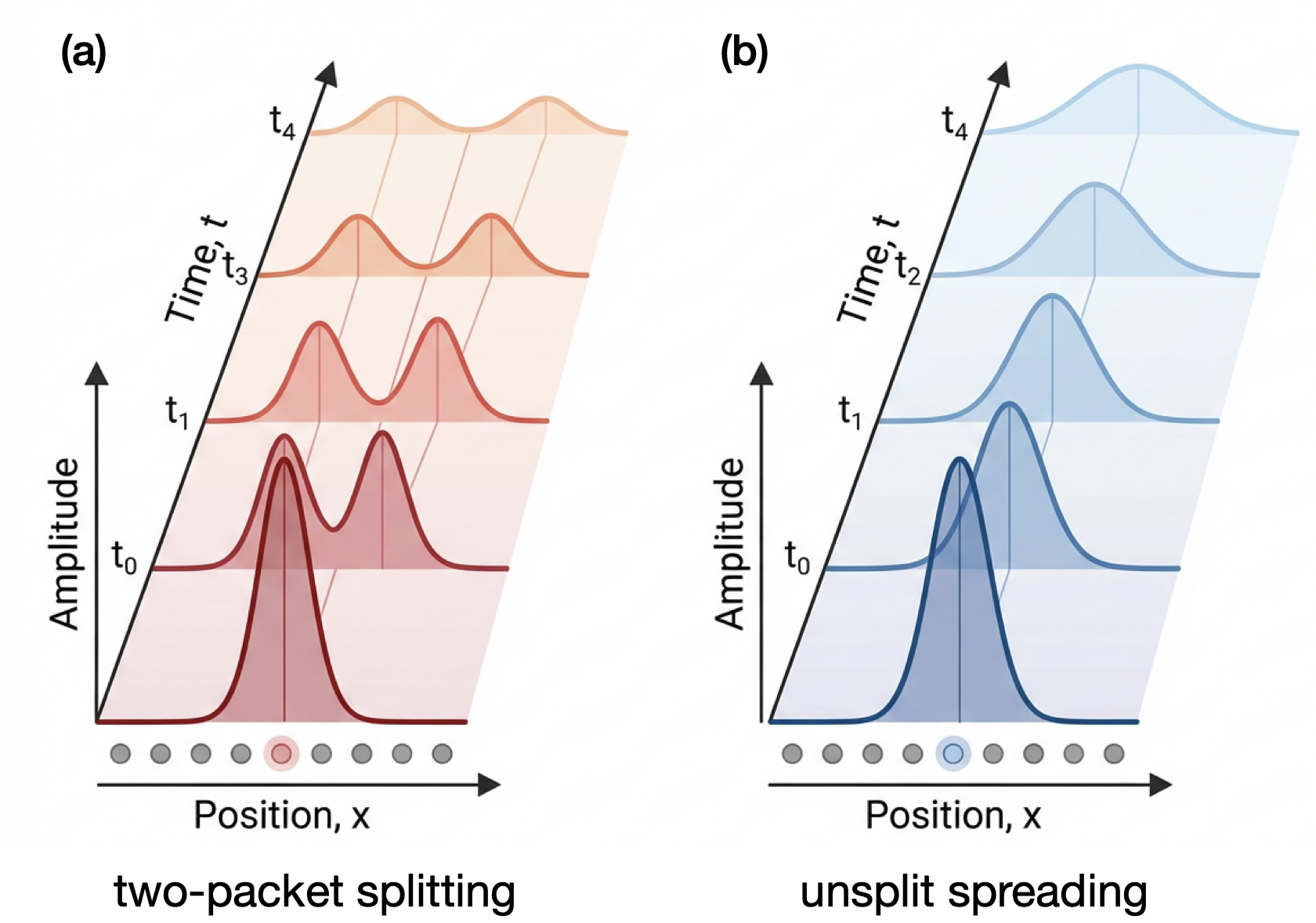}
    \caption{An initially localized excitation on a lattice exhibits two distinct diffusion behaviors: (a) it splits into two broadening wave packets that propagate in opposite directions; (b) it spreads without splitting. The solid curves represent the envelopes of the waveforms. As we will demonstrate, behavior (b) is a signature of singularities in the system’s dispersion relation arising from long-range interactions.}
    \label{fig:schematic}
\end{figure}

In this Letter, we provide a conceptually transparent answer.  
Consider a local excitation initialized at a single site of a 1D lattice.  
The excitation evolves into a time-dependent waveform.  
While the Lieb--Robinson bound constrains the speed of the propagating wavefront~\cite{Lieb:1972aa,Hauke:2013aa,Gong:2023aa,Chen:2019aa,Else:2020aa,Richerme:2014aa}, we focus on the overall shape of the wave packet.  
As illustrated in Fig.~\ref{fig:schematic}, two distinct dynamical patterns emerge: (a) \emph{two-packet splitting}, where the excitation separates into two counter-propagating wave packets, and (b) \emph{unsplit spreading}, where the waveform broadens without splitting.  
Pattern (b) resembles the textbook spreading of a Gaussian wave packet, yet we prove that it is impossible on a lattice whenever the band dispersion $\omega(k)$ is smooth.  
This no-go theorem admits a topological interpretation, and we extend it to 2D lattices by invoking the Gauss--Bonnet theorem.  
Consequently, the presence or absence of splitting serves as an observable signature of the non-smooth dispersion induced by long-range interactions.  
Indeed, although not explicitly noted, we identify unsplit spreading in Fig.~4(c) of Ref.~\cite{Jurcevic:2014aa}, where a 1D long-range XY model with $\alpha\approx 0.75$ was simulated using trapped ions.

In the following, we first prove that unsplit spreading is forbidden for smooth bands and then generalize the result to 2D lattices.  
Subsequently, we demonstrate unsplit spreading both in standard tight-binding models with $1/r^\alpha$ hoppings and in realistic models of light-mediated interactions.

\paragraph{No-go theorem for unsplit spreading.} 
We assume the lattice dispersion relation $\omega(k)$ is 
twice continuously differentiable (i.e., $\mathcal{C}^2$), 
so that its first and second derivatives are continuous.
The wavefunction of an excitation initialized at 
the site $x=0$ reads
\be\label{eq:xt}
\psi(x,t)=\int_{-\pi}^{\pi}\frac{dk}{2\pi} e^{-i\omega(k)t+ikx},
\ee
where the normalization is omitted. 
To evaluate Eq.~\eqref{eq:xt}, for given coordinate $x$ and time $t$,
we follow the standard stationary-phase approximation and
find the Bloch waves whose group velocity $v_g$ equals $x/t$:
\be\label{eq:vg}
v_g(k)\equiv\partial_k \omega(k)=x/t.
\ee
Equation~\eqref{eq:vg} may admit multiple solutions,
which we denote by $\tilde{k}_{i}$, indexed by $i$. 
Around each $\tilde{k}_{i}$, we expand
$\omega(k)$ to the second order and obtain
\begin{align}
\psi(x,t) & \approx \int_{-\pi}^{\pi}\frac{dk}{2\pi} 
\sum_{i} e^{-it\big[\omega(\tilde{k}_{i})
+\frac{1}{2} \partial_k^2 \omega(\tilde{k}_{i})(k-\tilde{k}_{i})^2 
\big]} \nonumber \\
& \approx \frac{1}{2\sqrt{\pi}}\sum_{i}e^{-i\omega(\tilde{k}_{i})t}
\big[t\partial_k^2 \omega(\tilde{k}_{i})\big]^{-1/2},
\label{eq:2nd}
\end{align}
where to obtain~\eqref{eq:2nd} we have extended the 
integral to the whole real axis. 
Although this is not quantitatively accurate (see the Supplemental Material~\cite{sp}
for a comparison with exact results), it clearly reveals the dependence on $\partial_k^2\omega(k)$. 
This is physically meaningful because
$1/\abs{\partial^2_k\omega(k)}$ is proportional to the
density of Bloch states sharing the same 
group velocity. Consequently,
regions where $\abs{\partial^2_k\omega(k)}$ is small tend
to exhibit enhanced wavefunction amplitudes, especially 
when $\partial^2_k\omega(k)=0$. To locate these peaks,
we first solve the equation
\be\label{eq:w2=0}
\partial_k^2 \omega(k)=0.
\ee
Denoting its solutions by $k^*_i$, the corresponding peak
positions are approximately
$x_i(t)=v_g(k_i^*)t$. 

The number of such peaks and
their motions thus determine the overall shape of the wavefunction.
Since $\omega(k)$ and its first two derivatives are continuous and 
periodic over the first Brillouin zone, we have
\be\label{eq:int}
\int_{-\pi}^{\pi}\frac{dk}{2\pi}\partial^n_k\omega(k)=0,\;\; \text{for}\; n=1, 2.
\ee
We ignore the case of flat bands. If $\partial^2_k\omega(k)>0$ is positive at some point,  
the condition~\eqref{eq:int} with $n=2$ implies it
must take negative values elsewhere. Combined with periodicity,
which requires $\partial^2_k\omega(k)$ to return to its initial
value after a full period, it is apparent that $\partial^2_k\omega(k)$ crosses zero
an even number of times, generally at least twice. Each
zero-crossing point corresponds to a pronounced peak of the waveform.
By definition, $k^*_i$ is where the group velocity obtains extremum.
These extrema must take both positive and negative values due to the
continuity and periodicity of $v_g(k)$. 
These correspond to right-moving and left-moving wave packets, respectively, giving rise to the split profile illustrated in Fig.~\ref{fig:schematic}(a).

To summarize, ``unsplit spreading'' is ruled out by mere smoothness and periodicity of $\omega(k)$. The most crucial step is 
that $\partial_k^2 \omega(k)$ has at least two zeros. 
This is actually a topological feature of smooth functions on
1D circle ($S^1$). Although elementary mathematics is sufficient for 1D, 
we find topological viewpoint essential for 2D, where we need to consider the zeros of
the determinant of 
Hessian $\mathcal{H}_{ij}(\vec{k}) \equiv \partial_{k_i} \partial_{k_j} \omega(\vec{k})$.

Given a non-flat dispersion, the graph $\Sigma = \{(\vec{k}, \omega(\vec{k}))\}$ defines a smooth surface in $\mathbb{R}^3$ 
generally diffeomorphic to the 2D torus $T^2$. The Gauss--Bonnet theorem 
specified on $\Sigma$ reads
\be
\frac{1}{2\pi }\int_{\Sigma} K(\vec{k}) \, dA = \chi(T^2) = 0,
\ee
where $K(\vec{k}) = \det(\mathcal{H}) / (1 + |\vec{v}_g|^2)^2$ is the Gaussian curvature of $\Sigma$, $\vec{v}_g(\vec{k}) = \nabla \omega(\vec{k})$
is the group velocity, $dA = \sqrt{1 + |\vec{v}_g|^2}\, d^2k$ is the induced area element on $\Sigma$, and $\chi(T^2)$ denotes the Euler characteristic of 2D torus.
Because the weight $(1 + |\vec{v}_g|^2)^{-3/2}$ is strictly positive, the 
vanishing integral forces $\det(\mathcal{H})$ to take both positive and negative 
values across the first Brillouin zone. 
Consequently, the zero set $\det(\mathcal{H}) = 0$ forms a 1D 
sub-manifold $\mathcal{S}$ (generically a collection of closed circles), separating 
regions of $\det(\mathcal{H})> 0$ and $\det(\mathcal{H}) < 0$.
Circles in $\mathcal{S}$ plays the same role of 
isolated points satisfying $\partial_k^2 \omega = 0$ in 1D. 
Along one circle in $\mathcal{S}$, one eigenvalue of the Hessian vanishes, while the other, denoted $\lambda(\vec{k})$, generally remains finite and influences the wavefunction amplitude. 
Since both $\vec{v}_g(\vec{k})$ and $\lambda(\vec{k})$ are smooth 
and periodic when restricted to $\mathcal{S}$, 
they generically lead to split spreading.

\paragraph{1D power-law interactions.} 
To circumvent the 1D version of our no-go theorem, 
which requires $\omega(k)$ to belong to
$\mathcal{C}^2(S^1)$,
we can only violate the assumption of smoothness.
Here we consider the standard models with power-law tunneling amplitudes 
$1/r^\alpha$. The corresponding
dispersion relation reads
\be\label{eq:wk}
\omega_{\alpha}(k)=\sum_{r=1}^{\infty} 
\frac{2}{r^\alpha} \cos(kr).
\ee 
For its derivatives, the differentiation $\partial_k$ and infinite summation 
do not commute in general. Nevertheless,
we may first differentiate the summand term by term and then examine the convergence of the resulting series. It turns out that
$\omega_{\alpha}(k)\notin \mathcal{C}^{k}(S^1)$ if $\alpha\leq k+1$. 
Thus, the cases of $\alpha=1,2$ and $3$ mark the threshold for
$\omega_\alpha(k)$, $\partial_k\omega_\alpha(k)$, and 
$\partial_k^2\omega_\alpha(k)$ to be discontinuous,
respectively. In the following, we discuss them one by one.

\begin{figure}[b]
    \centering
    \includegraphics[width=0.9\textwidth]{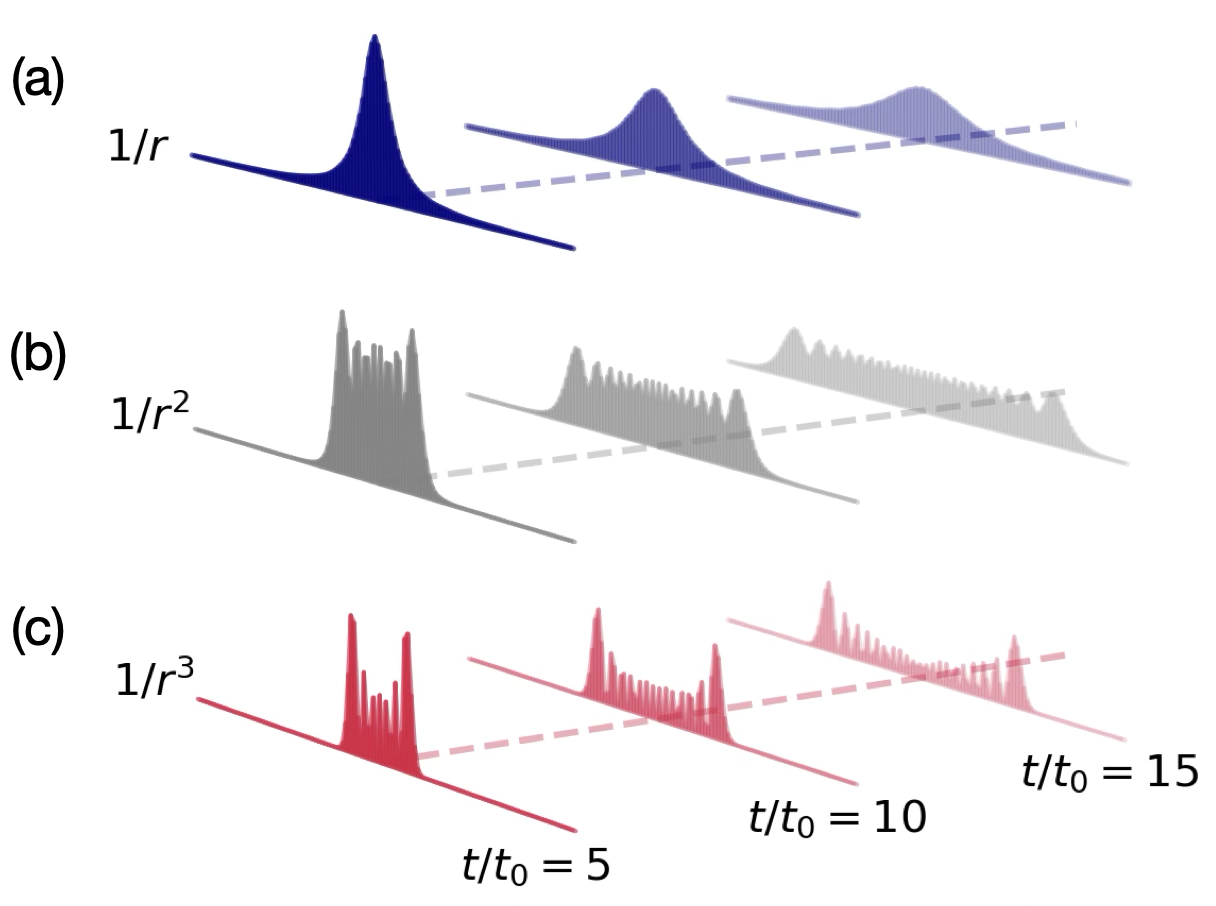}
    \caption{ Spreading of waveforms under long-range interactions decaying as $1/r^\alpha$, with (a)–(c) corresponding to $\alpha=1, 2$ and $3$, respectively. For each case, the waveforms at three different times are shown. Time is measured in units of $t_0=a^{\alpha}$ where $a$ is the lattice constant.}
    \label{fig:power}
\end{figure}

\paragraph{The case of $\alpha=1$.} This is the borderline case where 
$\omega_1(k)$ is not continuous. The series~\eqref{eq:wk} with $\alpha{=}1$ 
can be obtained by substituting 
$z{=}e^{ik}$ into the Taylor expansion of $F(z)=-\ln(1-z)-\ln(1-1/z)$ 
around $z=0$. $F(z)$ has a pole at $z=1$, hence, $\omega_1(k)$ diverges at
$k=0$. The first two derivatives are
\be
\ba
\partial_k \omega_1(k) & = -\mathrm{cot}(k/2), \\
\partial_k^2\omega_1(k) & =-1/[2\sin^2(k/2)]
\ea
\ee
Notably, $\partial_k^2\omega_1(k)<0$ (except for $k=0$) 
and its magnitude obtain a 
unique minimum at $k=\pi$. According to Eq.~\eqref{eq:2nd},
the waveform does not split. This is
confirmed by the waveform evolution shown in Fig.~\ref{fig:power}(a).

\paragraph{The case of $\alpha=2$.} This is the threshold
at which $\omega_\alpha(k)$ remains continuous but ceases to be
differentiable, i.e., $\omega_2(k)\notin \mathcal{C}^1(S^1)$. 
It is convenient to express $\omega_2(k)$
on the interval $[0,2\pi]$, where one finds
$\omega_2(k)=(3k^2-6\pi k+ 2\pi^2)/12$. Transforming back to 
$[-\pi,\pi]$, the derivatives read
\be\ba
\partial_k\omega_2(k) & =\begin{cases}
k+\pi, &-\pi\leq k<0; \\
k-\pi, & 0<k\leq \pi.
\end{cases} \\
\partial^2_k\omega_2(k) &=1,\quad k\neq 0.
\ea
\ee 
Thus, $\partial^2_k\omega_2(k)$ is a constant almost everywhere
except for $k=0$, where it is undefined due to the cusp in 
$\partial_k\omega_2(k)$. Now Eq.~\eqref{eq:2nd} suggests no obvious peaks. 
As shown in Fig.~\ref{fig:power}(b),
we observe an expanding rugged plateau, where the oscillations
on top of the plateau come from the interference 
caused by the phase factors of Eq.~\eqref{eq:2nd}.
We classify this profile as unsplit spreading.

\paragraph{The case of $\alpha=3$.} This is the threshold at
which $\omega_3(k)\notin \mathcal{C}^2(S^1)$. We find
$\partial_k^2\omega_3(k)=\ln[4\sin^2(k/2)]$, which diverges at $k=0$. 
Despite this singularity, the band structure now
supports two-packet splitting, as shown in Fig.~\ref{fig:power}(c). 
This is because $\partial_k^2 \omega_3(k)=0$
has two solutions at $k=\pm2\pi/3$. By parity symmetry,
$\omega_3(k)=\omega_3(-k)$,
the group velocities at $k=\pm2\pi/3$ have opposite signs,
implying two counter-propagating sub-packets. 
Notably, each sub-packet contains a train of subsidiary peaks.
This is typical for short-range models and 
is explained in the Supplemental Material~\cite{sp}.


\paragraph{Light-mediated interaction.} Now we turn to the 
light-mediated long-range  interactions in
1D and 2D \emph{subwavelength} atom arrays, where the lattice constant,
denoted by $a$, is shorter than resonant photon's 
wavelength. Before presenting the details, we note that 
in this system, band singularities will divide
the Brillouin zone into a dissipative sector and a coherent sector
(subradiant states). The latter, characterized by long lifetimes 
even in finite systems, effectively implementing a post-selection
from which unsplit spreading could be seen. This
offers a distinct pathway to circumvent the no-go theorem 
through open-system dynamics.

The light-mediated interaction is specified by
the dyadic Green's tensor of Maxwell's equations 
with the corresponding boundary conditions (can be engineered 
by photonic structures~\cite{Lodahl:2015aa,Yu:2019aa,Pennetta:2022aa,Tiranov:2023aa,Teifmmode:2024aa}). 
We assume translation symmetry and denote the dyadic tensor by 
$\bm{G}(\bm{r},\omega)$ with $\bm{r}$ and $\omega$ the coordinate and frequency augments, respectively.
Suppose the atoms have an excited state $\ket{e}$ and
a ground state $\ket{g}$. 
Two atoms can interact by exchanging photons through
the $\ket{e}{-}\ket{g}$ transition. In the Markov regime where 
the short-time dynamics~\cite{Zhang:2023aa} and the retardation effects
can be ignored~\cite{Sinha:2020aa}, 
the light-mediated interaction is described by
an effective atomic Hamiltonian $H=H_0+H_{\eff}$.
Therein, the free Hamiltonian $H_0$ reads 
$\sum_{i=1}(\omega_A-i\gamma_A/2)\sigma_i^\dagger \sigma_i$, 
where $\omega_A$ and $\gamma_A$ are transition frequency and
decay rate, respectively, of a single atom. 
The spin operators are defined by
$\sigma^\dagger=\ket{e}\bra{g}$ and $\sigma=\ket{g}\bra{e}$. The
interaction Hamiltonian is non-Hermitian~\cite{Lehmberg:1970aa,Lehmberg:1970ab,Dung2002}:
\be \label{eq:Heff}
H_{\eff}=-\mu_0\omega_A^2\sum_{i\neq j}
\bm{d}_i^{*}\cdot\bm{G}(\bm{r}_i-\bm{r}_j,\omega_A)\cdot\bm{d}_j\;
\sigma_i^{\dagger} \sigma_j,
\ee
where $\mu_0$ is the vacuum permeability, 
$\bm{r}_i$ and $\bm{d}_i$ denote the position and transition dipole 
of each atom, respectively. 
Below we consider 1D atomic chains coupled to either 1D
waveguide or free space, 
as well as 2D subwavelength atom arrays in free space.

\begin{figure}[b]
    \centering
    \includegraphics[width=\textwidth]{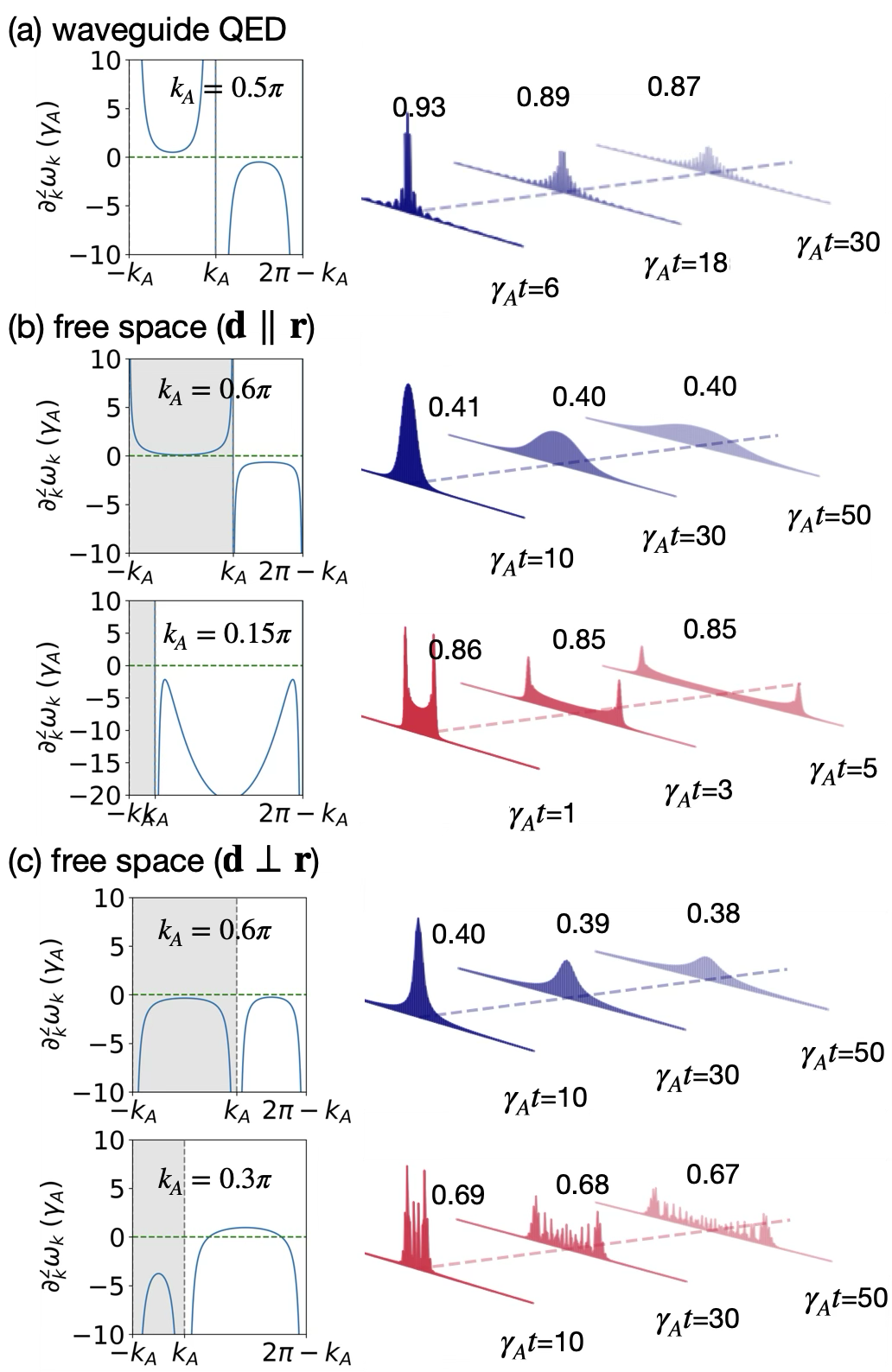}
    \caption{Left panels: $\partial_k^2\Re\omega(k)$, the second derivative of the real part of the dispersion relation. Right panels: snapshots of
    the waveform of an atomic excitation initialized at the center of a chain of
    751 atoms. Panels (a)–(c) correspond to (a) waveguide QED and (b, c) free space, with atomic dipole $\bm{d}$ polarized parallel and perpendicular to the chain, respectively.
    The first Brillouin Zone is shifted to $[-k_A, 2\pi-k_A]$ for visual convenience. The probability that the excitation survives decaying is indicated near each snapshot.}
    \label{fig:freespace_result}
\end{figure}

\paragraph{Waveguide QED.} An ideal 1D waveguide mediates an
``infinite''-range Hamiltonian~\cite{Chang2012,Chang:2018aa,Sheremet:2023aa}
\be 
H_{\mathrm{wg}}=-i\frac{\gamma_A}{2}\sum_{i\neq j} e^{-i k_A\abs{\bm{r}_i-\bm{r}_j}}
\sigma_i^\dagger \sigma_j,
\ee 
where $k_A=\omega_A/c$ and $c$ is the speed of light. 
As an non-Hermitian Hamiltonian, the corresponding dispersion relation 
is complex-valued. The
real part reads 
$\Re\omega_{\mathrm{wg}}(k)=(\gamma_A/4)\sum_{\varepsilon=\pm1}
\cot[(k_A+\varepsilon k)a/2]$~\cite{Zhang:2019aa}, and
the imaginary part reads 
$\Im\omega_{\mathrm{wg}}(k)=-i\gamma_A/(4a)\sum_{\varepsilon=\pm 1}\delta(k+\varepsilon k_A)$. 
It is not a coincidence that $\Re\omega_{\mathrm{wg}}$ and $\Im\omega_{\mathrm{wg}}$
are singular at the same locations,
because they are connected by the Kramers-Kronig relation.

In Fig.~\ref{fig:freespace_result}(a),  the left panel shows
$\partial_k^2\Re\omega_{\mathrm{wg}}(k)$ (see the plots of
$\omega_{\mathrm{wg}}(k)$ in the Supplemental Material~\cite{sp}). 
We observe two continuous branches
separated by the singular points $\pm k_A$, and each 
exhibits a nonzero minimum in absolute value.
Although such double minima signal splitting waveforms for $\omega(k)\in \mathcal{C}^2(S^1)$, here  
the group velocity vanishes at both extrema. Consequently, no spatial
splitting occurs, as confirmed by the three snapshots of
the waveform at in the right panel of Fig.~\ref{fig:freespace_result}(a).
Parameters used in simulation are introduced in the caption.

\paragraph{Atoms in free space.}  The dyadic Green's tensor of free space reads
\be\label{eq:G}
\ba
\bm{G}_{\text{fs}}(\bm{r},\omega_A)=\frac{e^{ik_A r}}{4\pi k_A^2 r^3}
\big[ & (k_A^2 r^2  +ik_A r-1)\mathbb{I}_3 \\
+ &  (-k_A^2 r^2-3ik_A r+3)\hat{\bm{r}}\hat{\bm{r}} \big],
\ea
\ee
where $\bm{r}=r\hat{\bm{r}}$ and $\hat{\bm{r}}\cdot \hat{\bm{r}}=1$,
and $\mathbb{I}_3$ is the $3{\times}3$ identity matrix.
Substituting $\bm{G}_{\text{fs}}$ into Eq.~\eqref{eq:Heff}
yields effective interactions containing 
$1/r^\alpha$ terms with $\alpha=1,2$ and $3$. 
The conventional coherent $1/r^3$ dipole-dipole interaction is
obtained in the limit of $k_A a\rightarrow 0$.
Importantly, when $k_A<\pi/a$, Bloch states with
$\abs{k}>k_A$ do not match with free-space radiation modes simultaneously in both energy and momentum.
This means the imaginary part of the dispersion relation, 
$\Im\omega_{\mathrm{fs}}(k)$, vanishes exactly 
for $\abs{k}>k_A$ (the subradiant states)~\cite{Asenjo-Garcia2017,Zhang:2020ab}.

In Fig.~\ref{fig:freespace_result}(b) we consider atomic 
dipoles polarized parallel to the chain, where
Eq.~\eqref{eq:G} does not contain
the $1/r$ term. We choose two lattice parameters, $k_A=0.6\pi$ and
$0.15\pi$ (with $a=1$), and plot $\partial_k^2\Re\omega_{\mathrm{fs}}$ 
in the left panels, where the unshaded region corresponds to the 
subradiant states.
For $k_A=0.6\pi$, the subradiant branch of
$\partial_k^2\Re\omega_{\mathrm{fs}}(k)$ attains its unique minimum 
magnitude at $k=\pi$, leading to unsplit spreading 
depicted in the right panel. In contrast,
for $k_A=0.15\pi$, the subradiant branch of $\partial_k^2\Re\omega_{\mathrm{fs}}(k)$
exhibit two local minima in absolute value, resulting two-packet splitting.
In Fig.~\ref{fig:freespace_result}(c) the dipoles are polarized
perpendicular to the chain. Now the $1/r$ term appears in
Eq.~\eqref{eq:G} but the main features remain unchanged,
except that for $k_A=0.3\pi$,
a train of subsidiary peaks appears as in Fig.~\ref{fig:power}(c).

\begin{figure}[t]
    \centering
    \includegraphics[width=\textwidth]{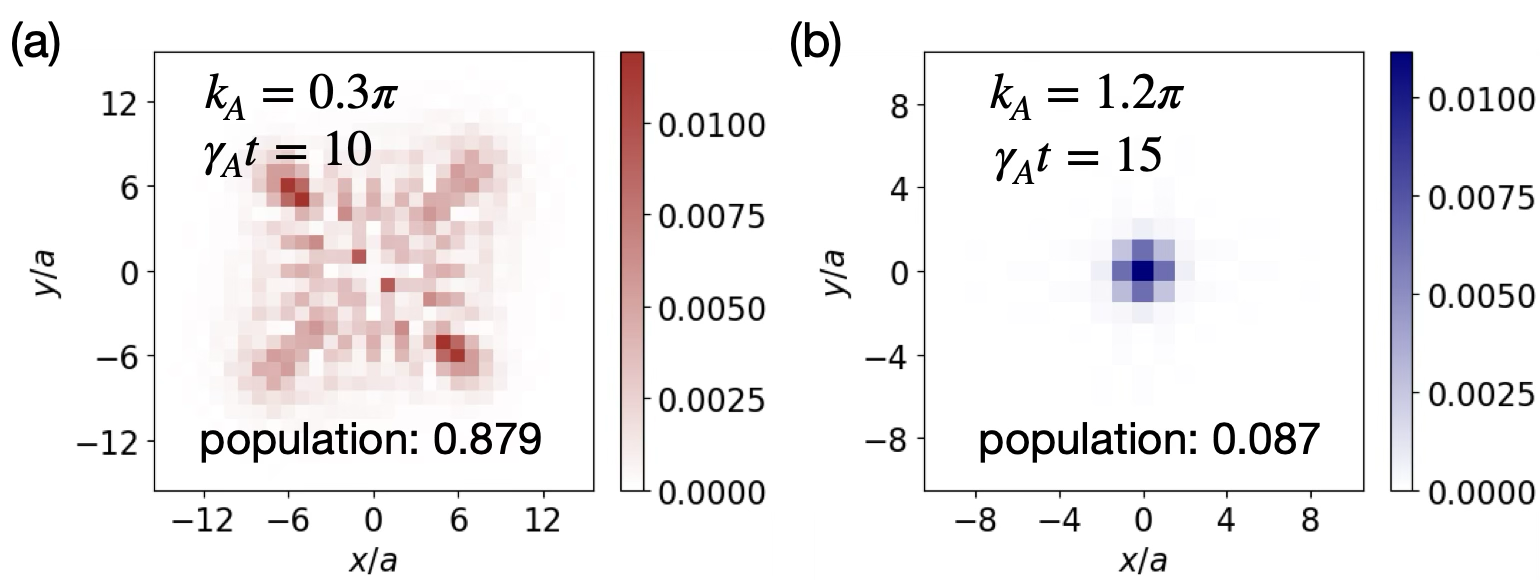}
    \caption{Waveforms of an excitation initialized in the center of a 2D square lattice ($93\times93$) with (a) $k_A =0.3\pi$; (b) $k_A =1.2\pi$. 
    The atomic dipoles are polarized along a fixed direction $\sin(\pi/12)/\sqrt{2}(\hat{\bm{x}}+\hat{\bm{y}})+
\cos(\pi/12)\hat{\bm{z}}$, where $\hat{\bm{x}}$ and $\hat{\bm{y}}$ 
are unit vectors along the lattice primitive directions and
$\hat{\bm{z}}$ is normal to the lattice plane. ``Population''
refers to the survival probability of the excited state against spontaneous emission.}
    \label{fig:2d_RDDI_result}
\end{figure}

\paragraph{2D subwavelength atom array.} Unfortunately, we do not see 
unsplit spreading in the standard $1/r^{\alpha}$ Hamiltonians on
square lattices~\cite{sp}.
While ``reverse engineering'' always works, i.e., we can 
in principle design a singular $\omega(\vec{k})$
and then reconstruct the corresponding lattice Hamiltonian, 
such models are typically highly
artificial. Instead, we focus on physically realizable 2D subwavelength
atom arrays in free space~\cite{Rui:2020aa}. We consider a finite square lattice 
atoms with size and dipole polarization specified in the caption of
Fig.~\ref{fig:2d_RDDI_result}. 
For $k_A =0.3\pi$ (with $a{=}1$), the waveform splits into multiple peaks,
as shown in Fig.~\ref{fig:2d_RDDI_result}(a). In contrast, for $k_A =1.2\pi$,
the excitation only diffuses weakly and remains unsplit,
as illustrated in Fig.~\ref{fig:2d_RDDI_result}(b). More plots
about 2D subwavelength atom arrays, and the discussion
about the consistency with the determinant
of the Hessian in the Supplemental Material~\cite{sp}, since 
the analysis is much more involved in 2D due to the $1/r$ terms.  

In the end, we note that the spreading of a localized 
excitation in 2D subwavelength atom array has been studied in 
Ref.~\cite{Ballantine:2020aa}. In this paper, it 
was shown that subradiance converts a localized initial
excitation into directional emission, which may find
applications in linking quantum networks. In the framework of
our theory, this effect can be understood as unsplit
spreading selected in orthogonal directions.

\paragraph{Conclusion.} In summary, we have proved
that unsplit wave-packet spreading is forbidden in lattices with smooth band structure,
which characterizes a fundamental gap between short- and long-range interactions.
This no-go theorem is actually a topological constraint
of continuous functions on the Brillouin zone, as demonstrated using 
the Gauss--Bonnet theorem for 2D
lattices. Long-range interactions circumvent this constraint by 
introducing singularities in the dispersion, enabling unsplit propagation. 
We also identify unsplit spreading in the subradiant
states of 1D and 2D subwavelength atom arrays, 
which is an open-system with natural post-selection.
Unsplit spreading, already present but overlooked in prior quantum simulation 
experiments~\cite{Jurcevic:2014aa}, 
provides a direct, observable signature of singular band structure. 
Our work invites exploration of interacting 
many-body generalizations. Additional interactions are unlikely to
remove the singular band structure generated by the free Hamiltonians.
Then, effects such as thermalization~\cite{Neyenhuis:2017aa,Sugimoto:2022aa} 
and entanglement growth~\cite{Koffel:2012aa,Schachenmayer:2013aa,Lerose:2020aa}
in long-range interacting systems,
may have features that can be traced back to singular dispersions.

\begin{acknowledgements}
Y.-X. Z. acknowledges the financial support from 
the National Natural Science Foundation of China (Grant No. 12375024),
the Innovation Program for Quantum Science and Technology (Grant No. 2023ZD0301100),  and the CAS Project for Young Scientists in Basic Research (Grant No. YSBR-100).
\end{acknowledgements}

\bibliography{long_subwavelength.bib}

\end{document}


\title{Supplemental Material to ``Singular Band Induced by Long-Range
Interactions Enables Unsplit Spreading of Localized Excitations}
\author{Jian-Feng Wu}
\author{Yi Huang}
\affiliation{Institute of Physics, Chinese Academy of Sciences, Beijing 100190, China}
\affiliation{School of Physical Sciences, University of Chinese Academy of Sciences, Beijing 100049, China}
\author{Yu-Xiang Zhang}
\email{iyxz@iphy.ac.cn}
\affiliation{Institute of Physics, Chinese Academy of Sciences, Beijing 100190, China}
\affiliation{School of Physical Sciences, University of Chinese Academy of Sciences, Beijing 100049, China}
\date{\today}

\begin{abstract}
    This Supplemental Material is arranged as follows.
    In Sec.~\ref{sp:sec_dispersion}, we plot the dispersion relations corresponding to
    the second derivatives shown in Fig.~3 of the main text.
    In Sec.~\ref{sp:sec_benchmark}, we benchmark the precision of the stationary-phase
    approximation used in Eq.~(3) of the main text.
    In Sec.~\ref{sp:sec_2D}, we present details about the calculation of 2D lattices, including the case of power-law decaying Hamiltonian and the 
    light-mediated interactions in 2D subwavelength atom arrays.
\end{abstract}

\maketitle

\section{Dispersion and Dissipation of Light-Mediated Interaction}
\label{sp:sec_dispersion}

\begin{figure}[htbp]
    \centering
    \includegraphics[width=\textwidth]{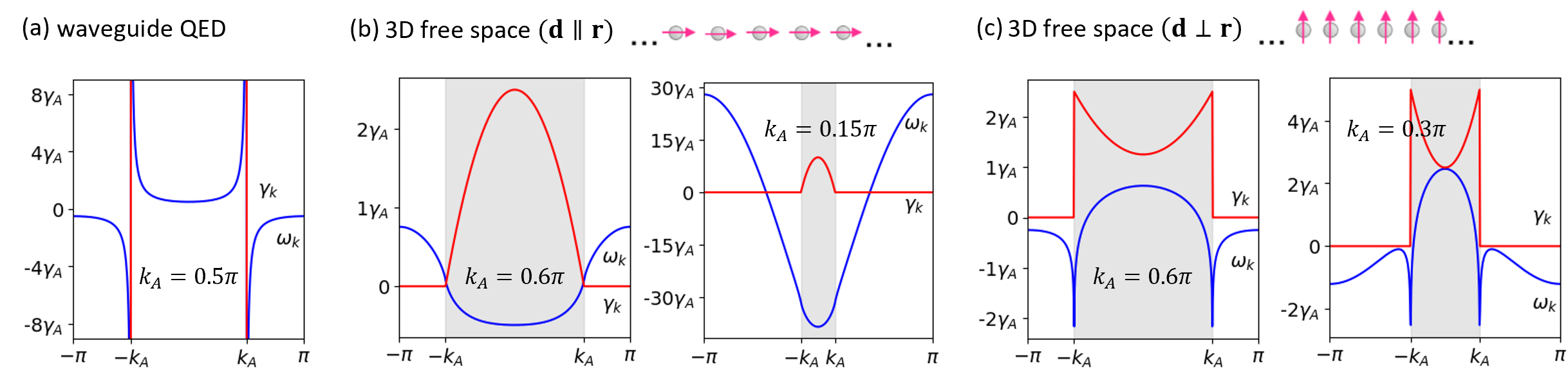}
    \caption{Real part (blue solid curve) and imaginary part (red solid curve) of the dispersion $\omega_k$ (in units of $\gamma_A$) for the setups of (a) waveguide QED; (b,c) atoms in free space with polarization $\bm{d}$ parallel and perpendicular to the chain, respectively. Values of $k_A$ are specified in the figures. We take lattice constant $a=1$.}\label{sp:fig_dispersion}
\end{figure}

In Fig.~(3) of the main text, we showed only the second derivatives of the real part of
the dispersion relations, $\partial_k^2\Re\omega(k)$. Here we plot the
corresponding complex-valued dispersion relations in Fig.~\ref{sp:fig_dispersion}. 

\section{Benchmarking the Stationary-Phase Approximation}
\label{sp:sec_benchmark}

\begin{figure}[htbp]
    \centering
    \includegraphics[width=\textwidth]{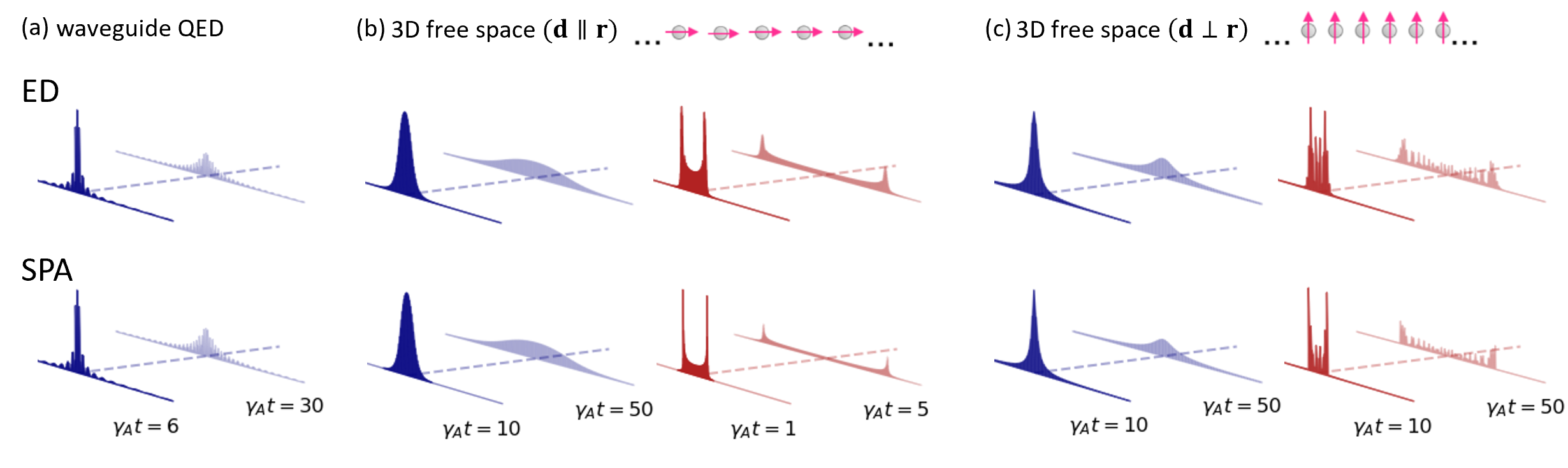}
    \caption{Comparing exact dynamics and stationary phase approximation for the setups of (a) waveguide QED; (b,c) atoms in the 3D free space with polarization $\bm{d}$ parallel and perpendicular to the chain, respectively. The waveforms of the first row are calculated from exact dynamics (ED), while the second row shows the results from stationary phase approximation (SPA). Values of $k_A$ are the same as those in the main text.}
    \label{fig:benchmarking_spa}
\end{figure}

In Eq.~(3) of the main text, we have approximated $\omega(k)$ by Taylor expansion
up to the 2nd order, and employed the Gaussian integral formula to
associated the dominant peaks of $\abs{\psi(x,t)}$ with the zeros of $\partial_k^2\omega(k)$,
at which Eq.~(3) of the main text diverges actually.
Thus, this approach cannot be quantitatively precise and we need to benchmark
to what extent it captures the waveforms.

We examine the stationary-phase approximation by applying it to all
the examples of 1D models
of light-mediated interactions studied in the main text. 
We apply Eq.~(3) of the main text to obtain $|\psi(x,t)|^2$ and normalize 
it, considering that this is an open-system evolution so that
the normalization should be the probability that no spontaneous emission occurs.
The results are shown in the lower panels of Fig.~\ref{fig:benchmarking_spa}. 
Comparing them with exact numerical simulation depicted in the upper panels
of Fig.~\ref{fig:benchmarking_spa},
we see that the stationary-phase approximation is not qualitatively good, 
but it does capture the profile.

\begin{figure}[htbp]
    \centering
    \includegraphics[width=0.5\textwidth]{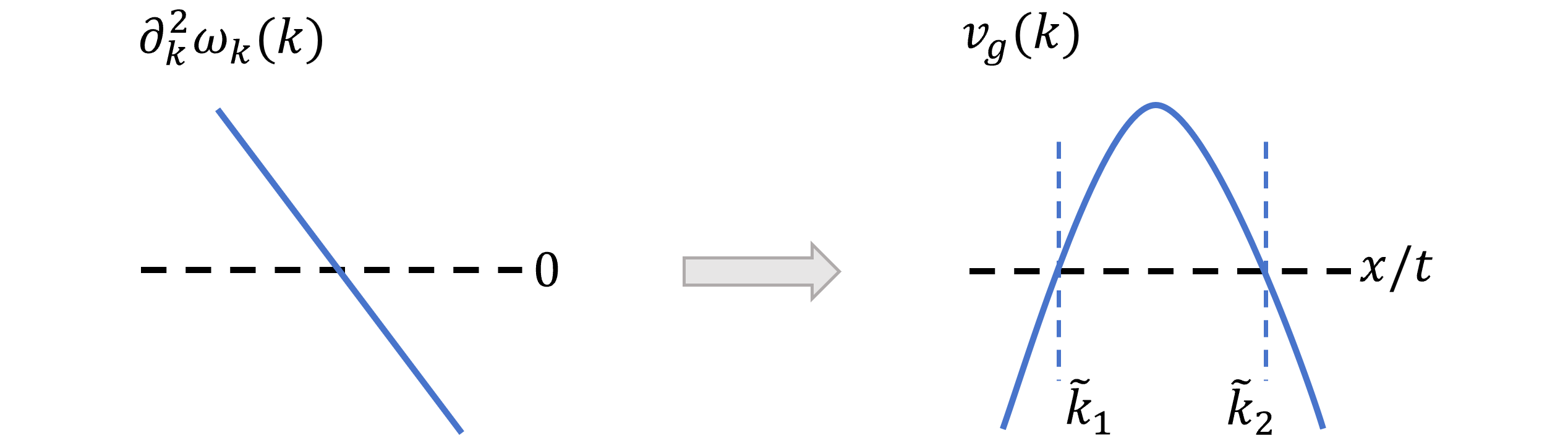}
    \caption{Schematic plot of the origin of train of subsidiary peaks. The non-monotonic behavior of $v_g(k)$ near $\partial^2_k\omega_k(k)=0$ shows multiple solutions of stationary points, which give rise to interference.}
    \label{fig:interference}
\end{figure}

Moreover, in Fig.~2(c) and Fig.~3(c) of the main text,
we observe that the splitting waveform has a train of subsidiary peaks, 
in contrast to the smooth profile displayed in Fig.~3(b) of the main text. 
This can be understood within the formalism of
stationary-phase approximation. Note that now we have zeros of $\partial^2_k \omega(k)$, 
as shown in the left panels of Fig.~3(c) of the main text. 
Around each zero point of $\partial^2_k \omega(k)$, the group velocity $v_g(k)$ is not monotonic with respect to $k$, which means that for a given $x/t$, there could be two solutions (see Fig.~\ref{fig:interference}). 
These two Bloch waves then interfere at $x=v_g t$, leading to the subsidiary peaks. 
Similar statements also apply to waveguide QED
depicted in Fig.~3(a) of the main text. Therein, the interference
comes from the Bloch modes belonging to different branches separated by $k=\pm k_A$.

\section{Spreading in 2D Lattices}
\label{sp:sec_2D}

\begin{figure}[htbp]
    \centering
    \includegraphics[width=0.7\textwidth]{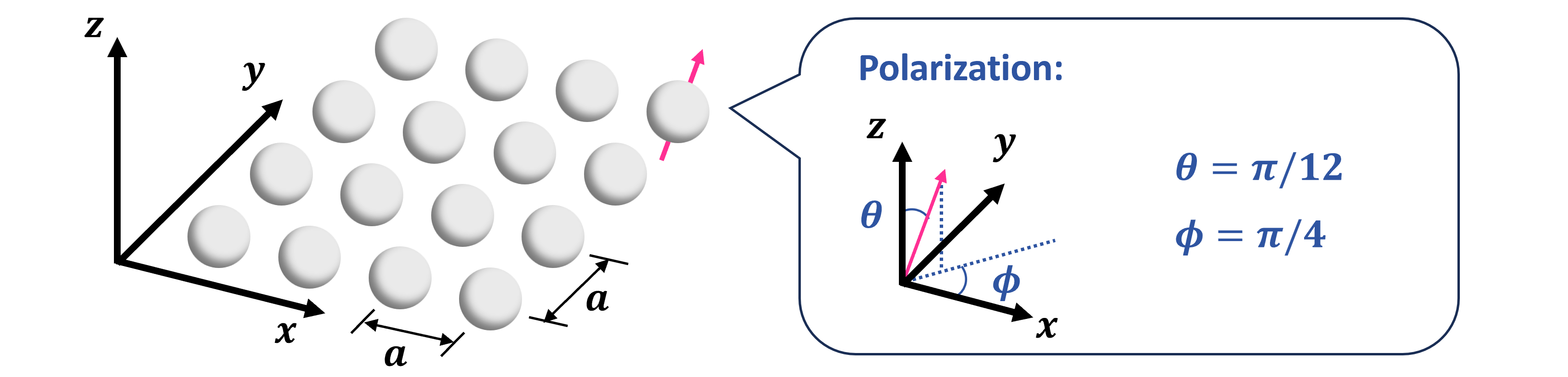}
    \caption{2D square lattice. Atoms are arranged in the $x-y$ plane and the polarization is parameterized by spherical coordinate $(\theta,\phi)$. The definitions of these angles are shown in the figure.
    }\label{sp:fig_2Darray}
\end{figure}

\subsection{Power-law hopping $(\alpha=2)$}

Here we show the spreading of a local excitation on 2D square lattice with power-law hopping $H = \sum_{i,j}J(r_{i,j})\sigma^\dagger_i\sigma_j  = \sum_{i,j} 1/r_{i,j}^{\alpha}\sigma^\dagger_i\sigma_j$ when $\alpha=2$. 
This is the threshold where we do not have a pure point spectrum but
still gain singularities in the dispersion relation.

\begin{figure}[htbp]
    \centering
    \includegraphics[width=0.7\textwidth]{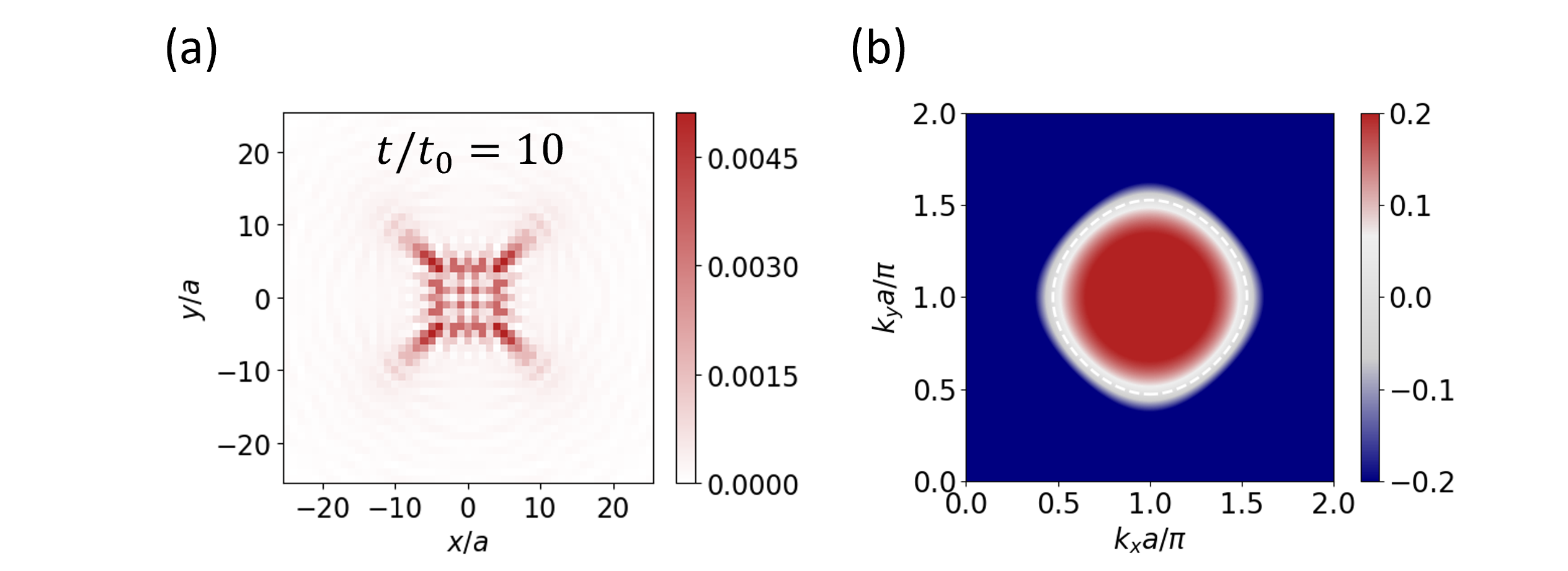}
    \caption{Waveform and $\det(\mathcal{H})$ (in units of $1/a^6$ where $a$ is the lattice constant) of 2D power-law model: (a) Exact dynamics on a $93\times93$ square lattice. Time is measured in units of $t_0=a^{2}$; (b) Determinant of Hessian calculated from dispersion (the discrete Fourier transformation of $J(r_{ij})$) using finite differences. The white dashed line denotes $\det(\mathcal{H})=0$.}
    \label{fig:2d_powerlaw}
\end{figure}

In contrast to the corresponding 1D case where $\alpha=1$, the waveform now splits into multiple peaks as shown in Fig.~\ref{fig:2d_powerlaw}. 
We further check the determinant of Hessian [$\mathcal{H}(\vec{k})$] 
using finite differences, and find that there is a closed loop that satisfies $\det(\mathcal{H})=0$. As analyzed in the main text, 
this is where the splitting waveform results from.

\subsection{2D subwavelength atom arrays}

Firstly, in Fig.~\ref{sp:fig_2Darray} we illustrate the setup and the direction
of the polarization of the atomic transition dipoles used in the main text.

Next, we discuss the corresponding $\det(\mathcal{H})$ of 
2D subwavelength atom arrays. In the main text, we consider a finite array
so that all the simulations are straightforward and meaningful. Here,
however, it becomes complicated 
to obtain converge dispersion relation due to the presence of $1/r$ terms.
As noted in the main text, it is not guaranteed that differentiation commutes with
infinite summation. 
In the literature of quantum optics, people usually ``regularize''
the Hamiltonians in order to circumvent the divergence.

\begin{figure}[htbp]
    \centering
    \includegraphics[width=\textwidth]{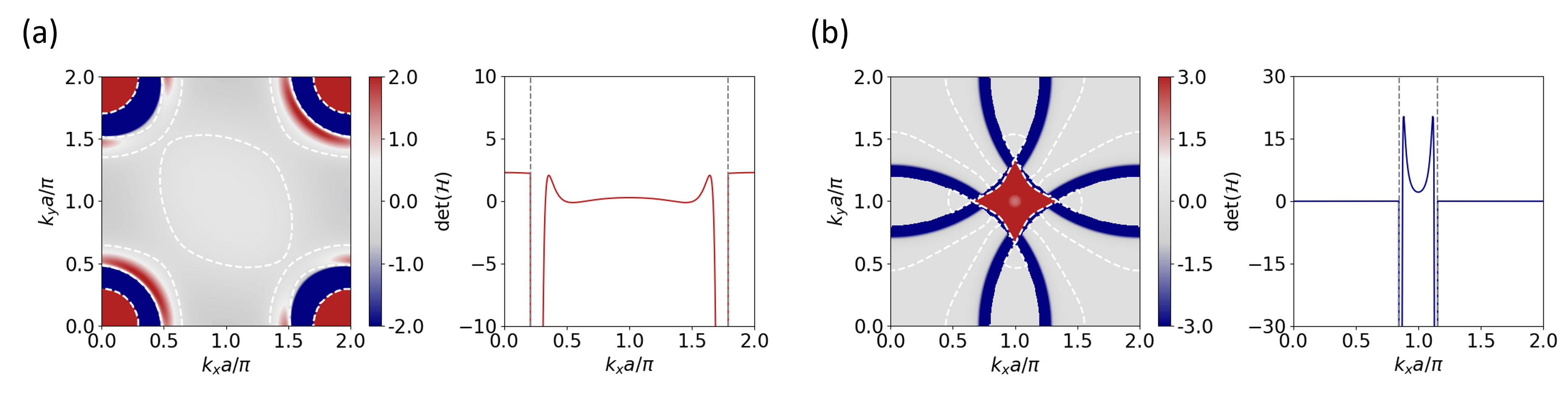}
    \caption{$\det(\mathcal{H})$ (in units of $\gamma_A/a^4$)  of 2D square lattice ($93\times93$) models calculated from regularized dispersion ($a_{\text{ho}}/a=0.1$) with (a) $k_A =0.3\pi$; (b) $k_A =1.2\pi$. Atoms arrangement and atomic polarization are the same as those in the main text. The white dashed lines on the left panels of (a)(b) denote $\det(\mathcal{H})=0$. On the right panels, determinants are plotted along $k_x=k_y$. Singularities at $\abs{\vec{k}}=k_A$ are marked by grey dashed lines.}
    \label{fig:2d_RDDI_detH}
\end{figure}

\begin{figure}[htbp]
    \centering
    \includegraphics[width=0.9\textwidth]{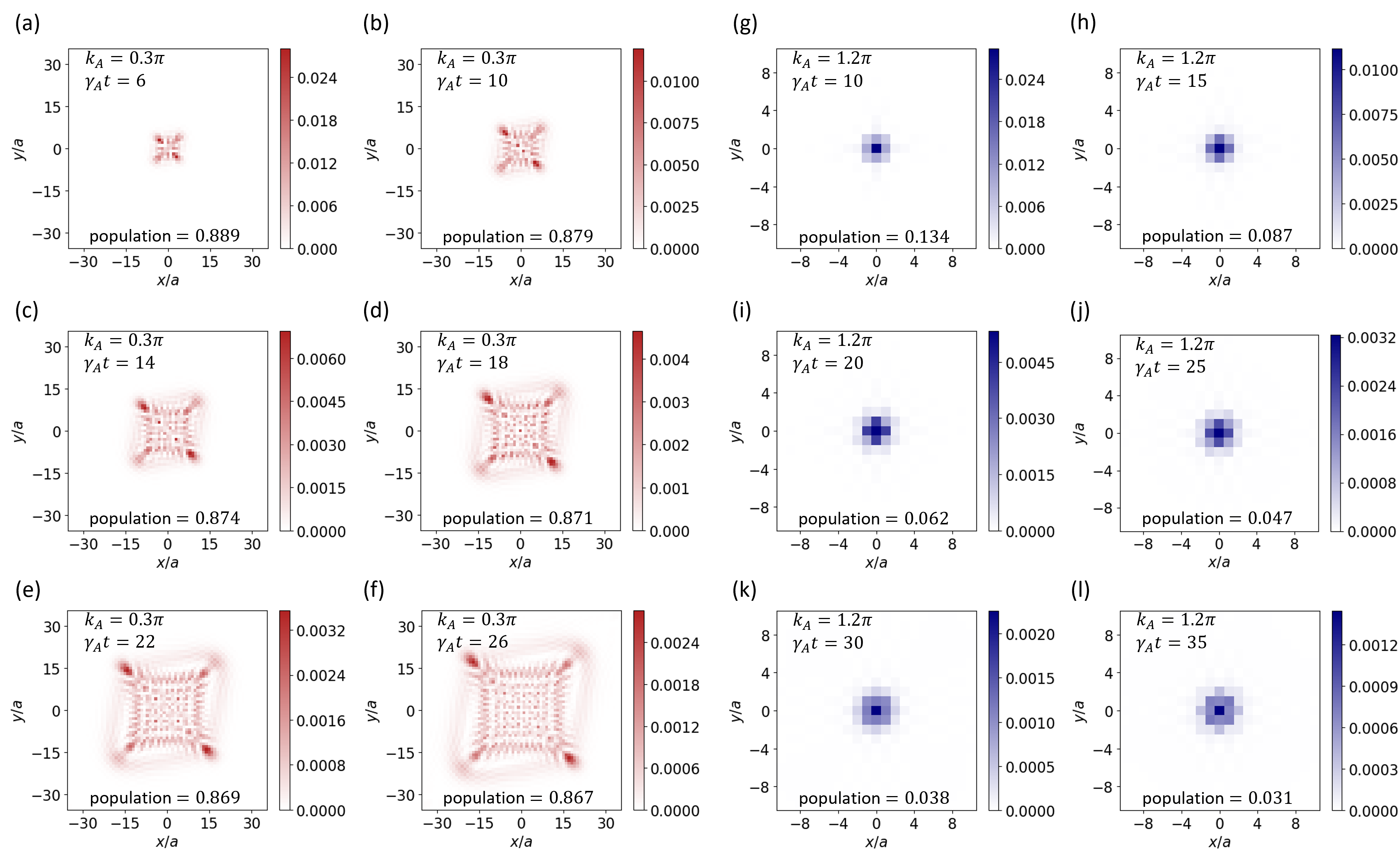}
    \caption{Waveforms at different $\gamma_A t$ on a 2D square lattice ($93\times93$) with (a-f) $k_A =0.3\pi$; (b-l) $k_A =1.2\pi$. Atoms arrangement and atomic polarization are the same as those in the main text. 
    }\label{sp:fig_2dmore}
\end{figure}

To proceed, we write the dispersion relation (real part) in the
form of summation over reciprocal space:
\be
\Re \omega_{\text{fs}}(\bm{k})/\gamma_A = -\frac{3\pi}{k_A}\Re \sum_{\bm{r}_{n} \neq 0} e^{i\bm{k} \cdot \bm{r}_{n}} \bm{G}_{\text{fs}}(\bm{r}_{n},\omega_A) = -\frac{3\pi}{k_A}\Re \left( \frac{1}{a_x a_y} \sum_{\mathcal{G} } \bm{g}(\mathcal{G} - \bm{k},\omega_A) - \bm{G}_{\text{fs}}(\bm{0},\omega_A) \right),
\ee
where $a_x,a_y$ are separations along $x,y$ axis respectively, $\mathcal{G}$ is the reciprocal lattice vector and $\bm{g}$ is defined as (differs from the definition in Ref.~\cite{Perczel:2017aa} by a sign due to the opposite sign convention for $\bm{G}_{\text{fs}}(\bm{r},\omega_A)$)
\be
\bm{g}_{\alpha\beta}(\bm{p},\omega_A) = \int \frac{dp_{z}}{2\pi} \frac{1}{k_{A}^{2}} \frac{k_{A}^{2} \delta_{\alpha\beta} - p_{\alpha} p_{\beta}}{p^{2} - k_{A}^{2}}.
\ee
Note that both parts on the right hand side diverge. 
Following the treatment of Ref.~\cite{Perczel:2017aa}, we assume the position
of each atom is not fixed in space but has a wavefunction with a finite spatial
uncertainty characterized by a length scale $a_{\text{ho}}$. It serves as
a regularization parameter with which the formula becomes
\be
\Re \omega_{\text{fs}}(\bm{k})/\gamma_A = -\frac{3\pi}{k_A}\Re \left( e^{k_A^2 a_{\text{ho}}^2/2} \frac{1}{a_x a_y} \sum_{\mathcal{G} } \bm{g}^*(\mathcal{G} - \bm{k},\omega_A) - \bm{G}^*_{\text{fs}}(\bm{0},\omega_A) \right).
\ee
The regularized $g^*$ reads
$$
\ba
g_{xx}^*(\bm{p},\omega_A) &= -(k_A^2 - p_x^2) \mathcal{I}_0, \\
g_{yy}^*(\bm{p},\omega_A) &= -(k_A^2 - p_y^2) \mathcal{I}_0, \\
g_{zz}^*(\bm{p},\omega_A) &= -(k_A^2 \mathcal{I}_0 - \mathcal{I}_2), \\
g_{xy}^*(\bm{p},\omega_A) &= g_{yx}^*(\bm{p},\omega_A) = p_x p_y  \mathcal{I}_0,\\
g_{xz}^*(\bm{p},\omega_A) & =g_{zx}^*(\bm{p},\omega_A)=g_{yz}^*(\bm{p},\omega_A)=g_{zy}^*(\bm{p},\omega_A)=0,
\ea
$$
where
$$
\ba
{\mathcal{I}}_{0}(\bm{p},\omega_A)&= {\mathcal{C}}\frac{\pi e^{-a_{\mathrm{ho}}^{2}\Lambda^{2}/2}}{\Lambda}\left[-i+\mathrm{erfi}\left(\frac{a_{\mathrm{ho}}\Lambda}{\sqrt{2}}\right)\right],\\
{\mathcal{I}}_{2}(\bm{p},\omega_A)&=\Lambda^{2}{\mathcal{I}}_{0}-{\mathcal{C}}\frac{\sqrt{2\pi}}{a_{\mathrm{ho}}},\\
\mathcal{C}(\bm{p},\omega_A) &= \frac{1}{2\pi k_A^2} e^{-a_{\mathrm{ho}}^2 p^2 / 2},\\
\Lambda(\bm{p},\omega_A) &= \sqrt{k_A^2 - p^2} .
\ea
$$
Following Ref.~\cite{Perczel:2017aa,Fernandez:2022aa}, we set $a_{\text{ho}}/a=0.1$ 
[we have examined that different choices of small $a_{\text{ho}}$ can change
the dispersions quickly, but the derivatives are relatively less sensitive 
to $a_{\text{ho}}$ in regions where $\det(\mathcal{H})$ does not diverge].  
and calculate the derivatives of $g^*$ to get $\det(\mathcal{H})$. Results are shown in Fig.~\ref{fig:2d_RDDI_detH}. 
For $k_A=0.3\pi$, there are closed loops of $\det(\mathcal{H})=0$ 
inside the subradiant zone, which give rise to the splitting waveforms. 
This is further supported by more snapshots of the
waveform evolution in Fig.~\ref{sp:fig_2dmore}(a).

For $k_A=1.2\pi$, while we see unsplit spreading in Fig.~4(b) of the main text, 
the regularized Hamiltonian still displays a loop of $\det(\mathcal{H})=0$ in 
the subradiant zone, as shown in Fig.~\ref{fig:2d_RDDI_detH}(b). However, by choosing $k_A=1.2\pi$, this loop
is adjacent to the diverging singularities at $\abs{\vec{k}}=k_A$,
which is also the boundary of the subradiant regime. This is
clearly demonstrated in cross-section depicted in the right panel of Fig.~\ref{fig:2d_RDDI_detH}.
It turns out that in finite systems, Bloch modes corresponding to this loop
still has a significant rate of spontaneous emission, i.e., a short lifetime.
We can then expect that the dynamics is dominated by the inner part of 
the subradiant zone.
which is free from zeros of $\det(\mathcal{H})$. In Fig.~\ref{sp:fig_2dmore}(b)
we show more snapshots of the waveform at different times. The plots
confirm unsplit spreading in this case, in contrast to the splitting
waveforms displayed in Fig.~\ref{sp:fig_2dmore}(a).

\begin{figure}[htbp]
    \centering
    \includegraphics[width=0.9\textwidth]{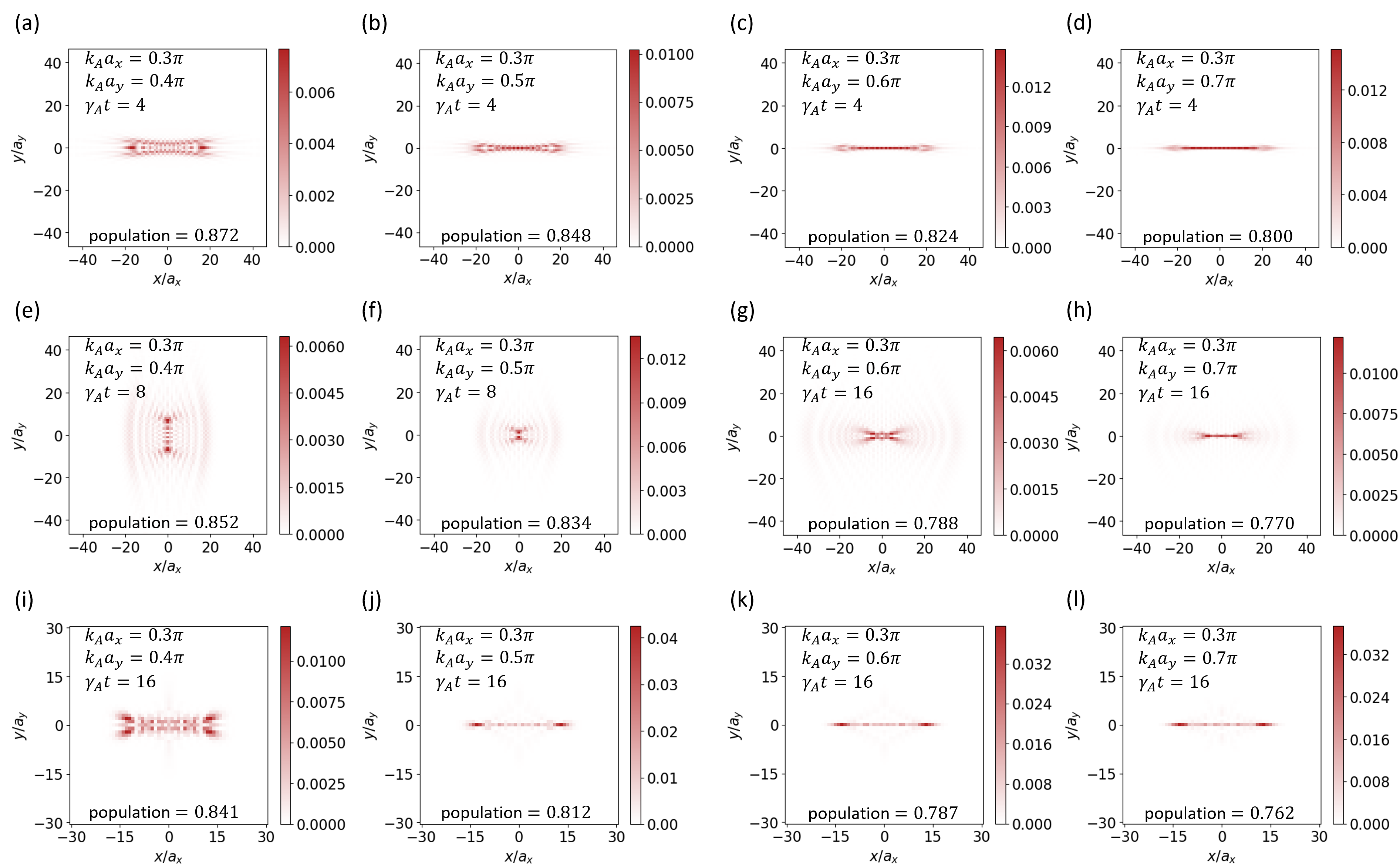}
    \caption{Waveforms on 2D rectangular lattices ($93\times93$) with polarization along (a-d) $x$ axis; (e-h) $y$ axis; (i-l) $z$ axis. 
    }\label{sp:fig_unsymmetry}
\end{figure}

Moreover, we can tune the parameters of the 2D subwavelength atom arrays to
see splitting in one direction, and unsplit spreading on the other direction.
This is illustrated for various lattice parameters in Fig.~\ref{sp:fig_unsymmetry}.

\bibliography{long_subwavelength.bib}